\documentstyle[11pt,newpasp,twoside,epsf]{article}
\def\edcomment#1{\iffalse\marginpar{\raggedright\sl#1\/}\else\relax\fi}
\reversemarginpar

\begin{document}
\title{Dust in Protoplanetary Disks}
 \author{E. I. Chiang}
\affil{UC Berkeley Astronomy, 601 Campbell Hall, Berkeley CA 94720}

\begin{abstract}
We critically examine the best lines of evidence for grain growth in
protoplanetary disks, based on modelling of observed spectral
energy distributions and images of T Tauri and Herbig Ae stars.
The data are consistent with millimeter-sized grains near
disk midplanes, and micron-sized grains near disk surfaces.
We review three channels by which grains can grow, including
direct condensation from the vapor phase, grain-grain collisional
sticking, and gravitational instability. The utility of dust
in identifying as yet unseen extrasolar planets is highlighted.
\end{abstract}

\section{Introduction}
By studying dust in disks surrounding pre-main-sequence stars,
we believe we glimpse the progenitors of planets.
In this review, we ask three questions: What is the observational
evidence for grain growth in circumstellar, presumably
protoplanetary disks? What is our theoretical understanding of
how grains grow from the submicron sizes that typify grains
in the diffuse interstellar medium (ISM) to the kilometer-sized
planetesimals that furnish the building blocks of planets?
And finally, how can we use dust to trace the presence
of extrasolar planets?

\section{Observational Evidence for Grain Growth}

\subsection{Models}

Interpretation of emission from grains in disks relies on models
for how radiation is transferred through disks. The simplest models
to consider are passive; the disk derives its luminosity by re-processing
incident light from the central star. Models of passive disks
in radiative and hydrostatic equilibrium are constructed by
Chiang \& Goldreich (1997, hereafter CG97) and Chiang et al.~(2001, hereafter
C01). The models assume that dust dominates the broadband
opacity of the disk from optical to millimeter wavelengths.
We examine in Figures 1 and 2, respectively, a schematic of how radiation
is transferred in externally irradiated disks,
and a sample spectral energy distribution (SED). We distinguish
two contributions to the emission above the Rayleigh-Jeans tail
of the stellar blackbody. At wavelengths, $\lambda$, of a few microns
to $\sim$60 $\mu$m, emission arises from the hot, ``superheated'' surface
layers of the disk. By definition, grains in these surface layers directly
intercept radiation from the central star; these naked grains are heated
to temperatures above that of blackbody if their sizes, $a$, are smaller
than the wavelengths characteristic of the re-emitted radiation.
For the emissivities of bare silicates and ice-coated silicates
employed by C01, surface grain temperatures are typically 3 $\times$
that of blackbody; for this reason, we refer to these grains as
``superheated.''\footnote{Not to be confused with the phenomenon
of temperature spiking of $a \la 0.01 \; \mu$m grains in the
ultraviolet radiation field of the diffuse ISM.} These hot grains radiate
their heat more or less isotropically. Half of their thermal emission
immediately escapes into space, where Earth-bound astronomers
perceive it to arise from an optically thin medium. The other half
is directed downwards into the deeper disk interior for further
reprocessing. The models of CG97 and C01 crudely treat the interior
as an isothermal (not necessarily optically thick) slab. The contribution
to the SED from the disk interior, which at a given stellocentric
distance, $r$, is cooler than the disk surface by a factor of $\sim$3,
is marked by the dashed line in Figure 2. It dominates at $\lambda \ga 100 \;
\mu$m. For more sophisticated treatments of radiative transfer that
spatially resolve the disk interior, see Calvet et al.~(1991),
D'Alessio et al.~(1998), and Dullemond, van Zadelhoff, \& Natta (2002).

\begin{figure}
\plotone{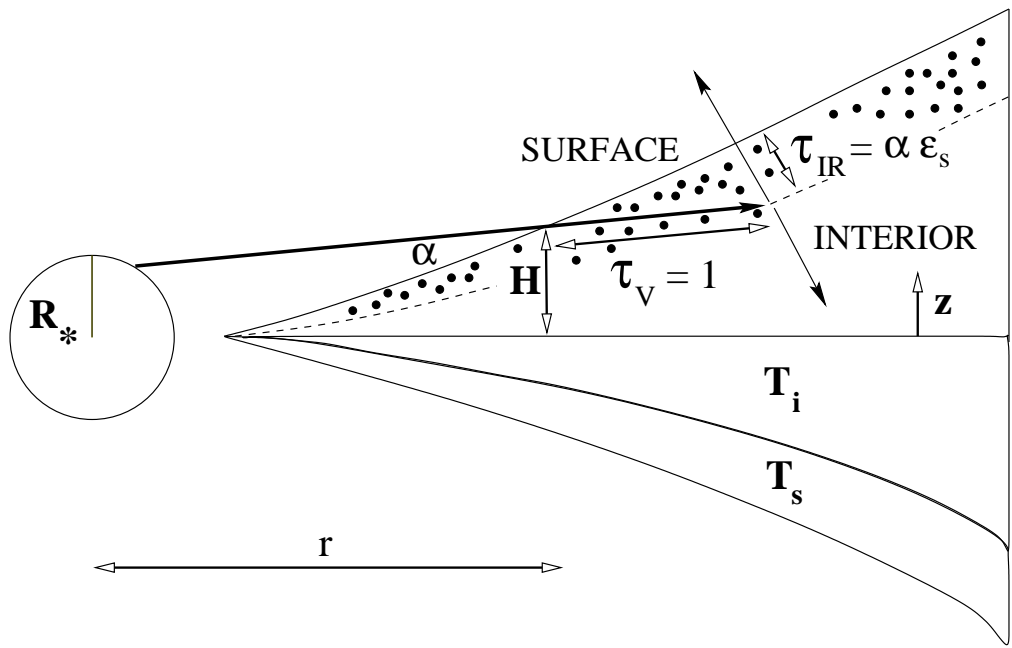}
\caption{Schematic of radiative transfer in passively reprocessing
circumstellar disks. The disk surface layer is defined as containing
grains that first intercept radiation from the central star. 
Surface grains radiate their heat isotropically; roughly half
of the reprocessed radiation immediately escapes into space,
while the other half is directed downwards into the disk interior
for further reprocessing. The height of the disk surface is geometrically
flared and enables the disk to reprocess substantially more starlight
at a fixed stellocentric distance than an actively accreting disk would.
For details and an explanation of the various variables,
see Chiang \& Goldreich (1997), from which this figure was taken.}
\end{figure}

\begin{figure}
\plotfiddle{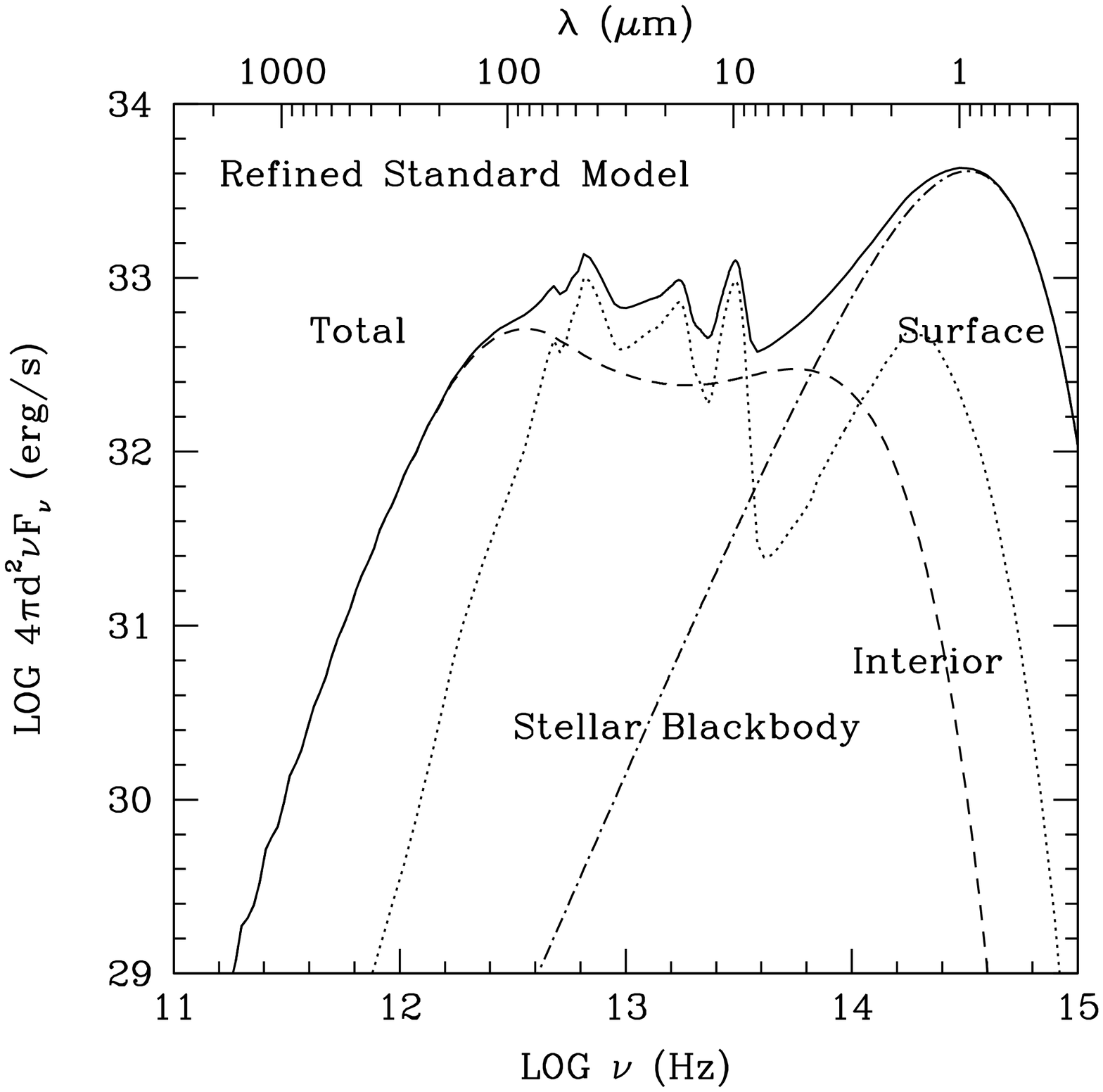}{225pt}{0}{60}{60}{-180}{-70}
\caption{Sample SED of a passive disk from Chiang et al.~(2001).
The central star contributes the blackbody spectrum peaked at 1 $\mu$m.
Disk emission can be divided into two parts, one from
the surface layers that dominate the flux at mid-infrared wavelengths,
and a second from the cooler disk interior that dominates at
$\lambda \ga 100 \; \mu$m. Surface grains are composed of amorphous olivine
and crystalline water ice; hence the solid-state
emission bands at $\lambda \sim 10$, 18, 45, and 60 $\mu$m.}
\end{figure}

The assumption of passivity is well justified since disks tend
to be geometrically flared; the pressure scale height of passively
reprocessing
disks scales as $h \propto r^{\gamma}$, where $\gamma > 1$.
To the extent that the height of the superheated surface layer, $H$,
follows $h$, the concavity of the disk
surface enables the disk to intercept more stellar energy at
a given $r$ than is locally released by accretion.\footnote{If dust and gas
are well-mixed in cosmic proportions, then $H/h$ varies from 5 to 4
across the entire disk, which spans $\sim$10$^2$ AU (C01).}
For this inequality
to hold, a rule of thumb is that $H$ must exceed the radius
of the central star. For typical
T Tauri parameters, passive reprocessing of starlight
dominates the local energy
budget at $r \ga 0.5$ AU or, equivalently, at $\lambda \ga 10 \; \mu$m.
The point is not that the disk is not actively accreting; most T Tauri
and Herbig Ae stars do evince ultraviolet (UV) excesses and veiling at optical
wavelengths that betray accretion in their immediate vicinity;
the point instead is that one would be hard pressed to tell whether
a disk is accreting based on the infrared (IR) SED alone. The insensitivity
of the infrared SED to the physics of disk accretion is a bane to
those who wish to understand the transport of mass and momentum
at large stellocentric distances, and a boon to those who wish
to model SEDs and infer disk temperatures, gas densities, and grain sizes.

By assuming that dust dominates the opacity in circumstellar
disks, we are saying that the forest of lines contributed by
circumstellar molecules such as H$_2$, CO, and H$_2$O cover
a small fraction of the continuum spectrum longward of $\sim$10 $\mu$m.
This is likely a fine assumption at $r \ga 1$ AU, where gas densities
are typically so low (a consequence of the weak tidal gravity field
of the star) that molecular lines should be thermally broadened
and therefore narrow (CG97). Closer to the star, pressure broadening
may render this assumption problematic. A careful accounting
of gas chemistry and opacity in the centralmost disk
should be performed. If recent explanations
(e.g., Dullemond, Dominik, \& Natta 2001) of the near-IR
excesses of Herbig Ae stars
in terms of central AU-sized cavities are to be believed,
such an accounting had better yield gas disks that are optically thin.

The main purpose of this subsection has been to provide a theoretical
backdrop against which to interpret observations of disks. From Figures
1 and 2, we recognize that thermal emission at mid-infrared wavelengths
probes the hot exterior skins of disks. Emission at millimeter
wavelengths probes conditions closer to the cooler disk midplane.
And scattered (optical-to-IR) light measurements necessarily
implicate disk surfaces. We proceed to examine critically
the best lines of evidence for grain growth that observations
and modelling currently offer.

\subsection{Grain Growth in the Disk Interior: Millimeter-wave Emission}

Claims of grain growth in T Tauri and Herbig Ae disks have been based
on the slope of the SED at millimeter wavelengths (e.g., Koerner, Chandler, \&
Sargent 1995); it is tempting to take the common observation
that $\delta \equiv d \log F_{\nu} / d \log \nu < 4$ to imply
that $\beta \equiv d \log \kappa_{\nu} / d \log \nu = \delta - 2 < 2$,
where $\kappa_{\nu}$ is the opacity and $\nu$ is the frequency
of observation. In the diffuse ISM where particle sizes do not approach
$\sim$1 mm, $\beta$ takes its value in the Rayleigh limit of 2.
Values smaller than 2 are held as evidence for grain growth
to millimeter sizes; in the geometric optics limit,
$\beta \approx 0$.

The problem with such reasoning is that it assumes the disk is optically
thin at wavelengths where the spectral slope is measured. Emission from
an optically thick medium would give $d \log F_{\nu} / d \log \nu = 2$
and a spurious inference that $\beta \approx 0$. Such issues
were appreciated by early and ground-breaking millimeter-wave studies
of pre-main-sequence stars (Beckwith et al.~1990; Beckwith \& Sargent 1991);
the review by Beckwith, Henning, \& Nakagawa (2000) carefully
outlines the assumptions and caveats behind conclusions based purely
on measured spectral slopes.
To address the concern of contamination by emission from optically thick
media, we must turn to detailed modelling of disks. Chiang et al.~(2001) fit
the IR-to-mm wavelength SEDs of 4 Herbig Ae stars and 1 T Tauri star
using a more sophisticated version of the 2-layer model of CG97.\footnote{In
particular, they correct a faulty estimate of CG97 that $\kappa_{\rm V}$,
the opacity of the dust-gas mixture at visible wavelengths, is of
order 400 cm$^2$/g. A more realistic estimate is closer to 3 cm$^2$/g.}
They account for grain size distributions and employ laboratory-measured
optical constants of silicates and water ice. In every pre-main-sequence
system, the mm-wave
SED is consistent with grain size distributions for which the mass
in concentrated in mm-sized particles. Grains are modelled
as (possibly ice-mantled) spheres in C01; this simplification is
not expected to be particularly restrictive; Henning \& Stognienko (1996)
compute opacities of fluffy, fractal aggregates and find values similar
to those of compact spheres. Unless grains are dominated by
highly conductive materials such as iron, grain shape is not expected
to alter opacities by more than factors of a few.

Despite this remarkable news that pre-main-sequence circumstellar disks
may be awash in sand-sized particles, these detailed modelling efforts
serve also to accentuate the severe degeneracies involved in fitting SEDs.
Figure 3, taken from C01, illustrates this degeneracy; the same dataset
can be fitted with either a small-mass disk containing large
grains or a large-mass disk containing small grains. The degeneracy
reflects the fact that the flux, $F_{\nu}$, from an optically thin
medium scales as its optical depth, $\tau_{\nu} = \Sigma \kappa_{\nu}$,
where $\Sigma$ is the disk surface density. One measurement of
$F_{\nu} \leftrightarrow \tau_{\nu}$ cannot break the degeneracy
between $\Sigma$ (disk mass) and $\kappa_{\nu}$
(grain size).\footnote{In principle, further uncertainty
exists in determining whether the observed emission arises
from a circumstellar disk at all; SEDs of spherical dusty envelopes
can be tuned to match SEDs of disks. This degeneracy in the
spatial distribution of dust can be broken by imaging.
In this review, all systems for which imaging data exist, either at optical
or millimeter wavelengths, are seen to be circumstellar disks.}

\begin{figure}
\plotfiddle{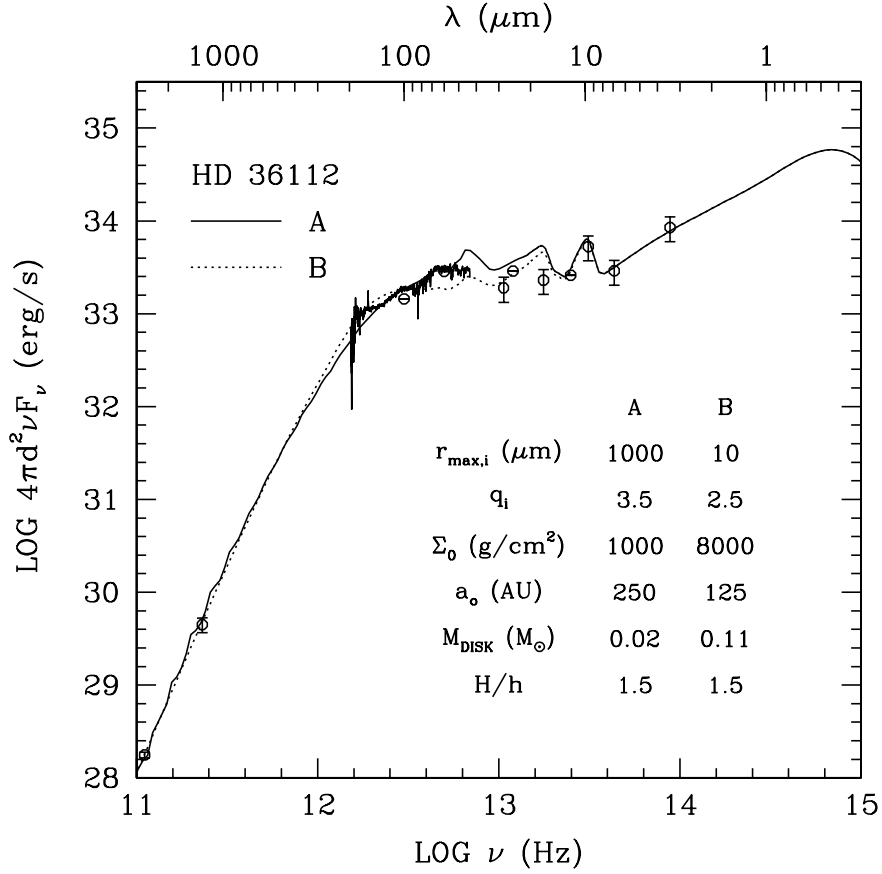}{300pt}{0}{60}{60}{-180}{-60}
\caption{Either a small-mass disk filled with large grains (model A)
and or a large-mass disk filled with small grains (model B) can fit
the SED of this Herbig Ae star. The disk mass (in gas and dust, mixed
in cosmic proportions) is given by $M_{\rm DISK}$, the maximum size
grain in the disk interior is given by $r_{\rm max, i}$, and
$q_i$ is the power-law index for the differential size distribution
of grains in the disk interior. For other symbols and a more in-depth
discussion, see Chiang et al.~(2001), from which this figure
was taken.}
\end{figure}

We are aware of one pre-main-sequence system for which this degeneracy
is tantalizingly close to being broken. That system is TW Hydra, a
8 Myr-old T Tauri star that sports a beautiful, face-on circumstellar
disk and is a mere 56 pc away. Weinberger et al.~(2002)
and Calvet et al.~(2002) both model the SED and find
the grain mass distribution to be weighted towards centimeter-sized
grains. Calvet et al.~(2002) proceed further to show that their
SED-based solution for the disk is consistent with the spatially
resolved mm-wave continuum images of the TW Hya disk. What is particularly
intriguing about these large grain-size solutions is that, if one
further assumes that gas and dust are mixed in cosmic proportions
of $\Sigma_{\rm gas}/\Sigma_{\rm dust} \sim 10^2$, then the
Toomre $Q$-parameter of this disk is near unity. Toomre's $Q$
measures the susceptibility of the disk to growth of perturbations
by self-gravity. It scales as $Q \propto 1/\Sigma$; values
far above unity imply stability, while values far below unity
imply violent instability on the local orbital timescale.
Discounting the large grain-size solution in favor of smaller grains
forces $\kappa_{\rm mm}$ to decrease and $\Sigma_{\rm gas}$ to increase,
such that $Q$ falls below unity. This is a recipe for violent
gravitational instability that we regard as unpalatable (see also
Johnson \& Gammie 2003). However, this argument
hinges on the assumption of a cosmic gas-to-dust ratio; as we
shall mention in \S3.3, solar metallicities may not characterize
disks that are sufficiently quiescent for dust to drift inwards relative
to gas.

\subsection{Grain Growth in the Surface Layer: Mid-IR Emission}
Silicate emission bands at $\lambda \sim 10 \; \mu$m and $\sim$20 $\mu$m,
routinely observed in the spectra of T Tauri and Herbig Ae stars,
imply that grains in hot disk surface layers are emitting in the Rayleigh
limit, $a \la \lambda / 2\pi \sim 2\; \mu$m. The emission band
at $\lambda \sim 10 \; \mu$m for TW Hya is spectrally well resolved
and constitutes a blend of two features, one centered at 9.6 $\mu$m
due to the Si-O stretching mode in glassy pyroxene, and another
at 11.2 $\mu$m due to the same vibrational mode in crystalline olivine
(Weinberger et al.~2002, and references therein). Though crystalline
silicates are occasionally evinced in spectra of pre-main-sequence
stars (Malfait et al.~1998), most systems exhibit predominantly
amorphous silicates. Natta, Meyer, \& Beckwith (2000) present
$\sim$10 $\mu$m emission band spectra of nine classical T Tauri
stars and reproduce them either with $a \approx 0.1 \; \mu$m grains
composed of glassy olivine (30\%) and glassy pyroxene (70\%),
or $a\approx 1\; \mu$m grains composed purely of glassy pyroxene.
Thus, a different degeneracy afflicts inferences of grain growth
in the optically thin surface layers: the degeneracy between
grain size and grain composition. Large grains exhibit broader
emission bands, but then so, too, do small grains having mixed mineralogies.

This degeneracy has recently been showing signs of strain. Van Boekel
et al.~(2003) report a trend, exhibited by more than 10 Herbig stars,
between the strength of the $\sim$10 $\mu$m feature and its width.
The broader the line, the lower is its ratio of peak height to continuum
flux. Variations in grain composition alone are argued to be incapable
of explaining this trend, since they would change line widths
but would affect line strengths less. By contrast, variations in grain size
would change both;
e.g., if the grain size becomes large enough that the geometric optics
limit is reached, then the line would disappear altogether.
Thus, the observed trend is most straightforwardly interpreted
as an evolutionary sequence of grain sizes;
some disks contain $a \approx 1\; \mu$m-sized grains in their
surface layers that produce both weaker and broader emission features
than are produced by the $a \approx 0.1\; \mu$m-sized
grains in other disks.

Another breakthrough observation is furnished by McCabe, Duch\^{e}ne,
\& Ghez (2003), who present the first spatially resolved image of
a T Tauri disk in {\it scattered} $\lambda \sim 11.8 \; \mu$m light.
Keck speckle interferometry proved particularly effective
for the opportune viewing geometry of HK Tau B, whose flared disk
is oriented at an inclination large enough to extinguish direct (blinding)
starlight but small enough to present nearly its full face as a scattering
surface. The spatial extent of the reflection nebulosity is $\sim$70 AU
(full width at half maximum),
far too large to be explained by thermal emission from surface
grains. The image is reproduced best by employing surface grains
having scattering asymmetry parameters of $g \approx 0.15$--0.83; smaller
$g$-values correspond to nearly isotropic emitters that produce
too much flux at large stellocentric distances, while larger
$g$-values generate a nearly unresolved point source.\footnote{The
$g$-value is the value of $\cos \theta$ averaged over the power
pattern of scattered light, where $\theta$ is the angle
relative to the direction of the incident beam. See, e.g.,
Bohren \& Huffman (1983), p.72.}
To the extent that surface grains can be modelled
using Mie theory, $0.15 \la g \la 0.83$ corresponds
to grain sizes $1.5 \la a (\mu\rm{m}) \la 3.2$ (McCabe et al.~2003).

These are not the only examples we can point to for evidence
of grain growth in disk surface layers. For example,
images of the edge-on disk in HH 30 at various near-infrared
wavelengths have led Cotera et al.~(2001) and Wood et al.~(2002)
to conclude that grains at least twice as large as those
typifying the maximum size grains in the ISM must be present
at altitude; if the unadulterated ISM mix were employed,
the variation in extinction
with wavelength would be too strong compared with observation.

To summarize these last two subsections, we have seen that the presence
of millimeter-sized grains in the dense midplanes of circumstellar
disks is certainly consistent with the millimeter-wave spectral
and imaging data, though to be fair, the existence of such large
grains is not unambiguously demanded. The presence in disk
surface layers of amorphous silicate grains having sizes of 0.1--2 $\mu$m,
with an occasional component of
crystalline silicates, is on more sure footing.
We proceed to describe in broad terms our theoretical
understanding of how grains grow.

\section{From Dust to Planetesimals: Theory}
We identify three channels for grain growth: accretion
from the vapor phase, grain-grain collisional sticking,
and gravitational instability.

\subsection{Condensation from Vapor}
Gas densities at the midplanes of typical models of protoplanetary disks are
so high that growth timescales for grains are extremely short
compared to the estimated lifetimes of disks. In the standard model
of the minimum-mass solar nebula (MMSN) as constructed by CG97,
the gas density (H, He, plus solar complement of metals) at the
midplane is $\rho_{\rm g} \sim 10^{-9} (r/{\rm AU})^{-39/14}$ g/cm$^{3}$.
A condensation nucleus that accretes metals from the vapor
phase increases its radius at the rate of

\begin{equation}
\frac{da}{dt} \sim \frac{Z \rho_g}{\rho_p} \; c_s \sim 1 \left( \frac{r}{\rm
AU} \right)^{-3} \; \rm{cm}/\rm{yr} \; ,
\end{equation}

\noindent independent of the size of the nucleus. Here $Z \sim 10^{-2}$
is the metallicity of the gas, $c_s \approx 0.1 \; (r/{\rm AU})^{-3/14}$ km/s
is the sound speed of condensing vapor, and $\rho_p \approx 2$ g/cm$^{3}$
is the internal density of the grain. Over the lifetime of the disk
of $\sim$10$^{7}$ yr, a condensation nucleus can grow to a maximum size of

\begin{equation}
a_{\rm max} \sim 10^7 \left( \frac{r}{\rm AU} \right)^{-3} \; {\rm cm} \; .
\end{equation}

\noindent Of course, before declaring victory in our effort to grow
grains, we must remember that the final size distribution of grains
that accrete from the vapor phase depends on the number of seed
condensation nuclei. The seed nuclei compete amongst each other
for a finite reservoir of metals. In our opinion, trying to
estimate the number of seed nuclei at the onset of grain growth
is a task not too far removed from trying to count the number of angels
on the head of a pin.

\subsection{Grain-Grain Sticking}
Numerous experiments, both laboratory-based and computer-simulated,
have been executed on aggregates of silica monomers (each monomer
having $a \approx 1 \; \mu$m)
to investigate the conditions under which such aggregates stick.
Blum \& Wurm (2000) find that similar-sized aggregates that
collide at relative velocities of less than $\sim$0.2 m/s
stick with little restructuring of aggregate bonds. Resultant
aggregates are highly porous, with large vacuum filling fractions
and a fractal geometry (see also Wurm \& Blum 1998).
At relative velocities
approaching $\sim$1 m/s, aggregates not only stick but also
compactify---the kinetic energy of the collision is diverted towards
rearranging bonds between monomers, and the aggregate is strengthened
as a consequence. Finally, relative velocities in excess of $\sim$1 m/s
shatter aggregates into their constituent monomers. These laboratory
results can be reconciled with numerical simulations by
Dominik \& Tielens (1997) if recently measured parameters
governing rolling friction and binding energy are used (Blum \& Wurm 2000).

Relative velocities between dust aggregates of less than $\sim$1 m/s
are certainly possible in protoplanetary disks (see Weidenschilling \& Cuzzi
1993). However, the actual velocity field of gas is neither strongly
observationally constrained nor understood, given our perennial ignorance
concerning sources of turbulence in protoplanetary disks. In a passive
(non-turbulent) nebula, we expect relative velocities between similarly
sized aggregates to grow with aggregate size, since larger particles
achieve greater terminal velocities as they gravitationally settle
towards the midplane. In the free molecular flow (Epstein) regime,
the vertical terminal velocity in disks is of order

\begin{equation}
v_{\rm term} \sim \sqrt{\frac{GM_{\ast}}{r^3}} \frac{\rho_p a}{\rho_g} \sim 4
\left( \frac{r}{\rm AU} \right)^{9/7} \left( \frac{a}{1 \; {\rm cm}} \right) \;
\rm{m/s} \; ,
\end{equation}

\noindent where $M_{\ast}$ is the mass of the central star
and $G$ is the gravitational constant
(see, e.g., CG97 or Youdin \& Chiang 2003). We have assumed here
that aggregates are sufficiently compact that they can be modelled
as spheres. Then we might expect grains to attain terminal
sizes of $\sim$$1 \;(r/{\rm AU})^{-9/7}$ cm, above which they would
be moving so quickly as to shatter each other. Our simple
estimate accords well with values cited in Blum \& Wurm (2000)
and Wurm, Blum, \& Colwell (2001).\footnote{See this latter
work and the related chapter in this book for a proposal
on how grain-grain sticking can exceed this maximum size.}

\subsection{Gravitational Instability}
Gravitational forces between dust grains become important when
their collective density exceeds the Roche density, $\sim$$M_{\ast}/r^3$.
The original expectation was that such densities would eventually
be achieved as dust settled towards the midplane into an ever
thinner layer (Goldreich \& Ward 1973). This hope was dashed
for a period of several years after it was realized that Kelvin-Helmholtz
turbulence generated within the particle layer prevented further settling
of dust before Roche densities were attained
(Cuzzi, Dobrovolskis, \& Champney 1993; Weidenschilling 1995).
The turbulence arises from vertical shear; dust-laden gas at the midplane
rotates at nearly the full Keplerian velocity, while relatively
dust-free gas residing above (and below) the midplane rotates more
slowly as a consequence of radial pressure gradients that (usually)
point outwards. Buffeted by the resultant Kelvin-Helmholtz
turbulence, dust fails to attain densities greater than that of gas,
falling short of the Roche density by 2 orders of magnitude.

Interest in gravitational instability has since been rekindled
by Sekiya (1998) and Youdin \& Shu (2002), who point out that
the aforementioned turbulence can be overcome in disks
having sufficiently super-solar metallicities. Turbulent
eddies in gas can entrain only a finite amount of dust;
the density of dust lifted to greater heights by turbulence
cannot exceed the gas density. Whatever excess dust is not
entrained must gravitationally precipitate out. Sekiya (1998)
quantifies these ideas by calculating the density of dust
as a function of height above the midplane in vertically
shearing (Cartesian) flows that are marginally Kelvin-Helmholtz
turbulent. Marginal turbulence means that the Richardson number
at every point in the flow is assigned its critical value
of 1/4 (see, e.g., Tritton 1988, p.350); dust is expected
to settle to near this critical state. Sekiya (1998) finds
that if the total, height-integrated dust-to-gas mass ratio
(metallicity) exceeds the solar value by more than a factor
of $\sim$10, then the dust density near the midplane
becomes formally infinite. Youdin \& Shu (2002) interpret the
appearance of this singularity as the onset of gravitational
instability.

How might super-solar metallicities be attained? Youdin \& Shu (2002)
discuss several metal enrichment processes, the simplest and most
robust of which is aerodynamically induced, radial drift of solids.
Solid particles encounter a headwind as they plow through sub-Keplerian
gas; the resultant friction causes particles to be dragged
radially inwards relative to gas. As particles migrate starward,
they can ``pile up'' and generate local enhancements in the surface
density of solids. Youdin \& Chiang (2003) model this accretion
process in detail, accounting not only for differences in the mean
flow velocities between particles and gas, but also for the turbulent
transport of angular momentum within the marginally Kelvin-Helmholtz
unstable particle sheet, and the dimunition of accretion velocities
as the particle density approaches the gas density. They conclude
that particle pile-ups are a robust outcome in protoplanetary
disks; in a few $\times$ $10^5$ yr---timescales shorter than
disk lifetimes of $10^7$ yr---metallicity enhancements of more than
a factor of $\sim$10 can occur at stellocentric distances of several AU.
A sample evolutionary sequence of surface density profiles
is showcased in Figure 4.

\begin{figure}
\plotfiddle{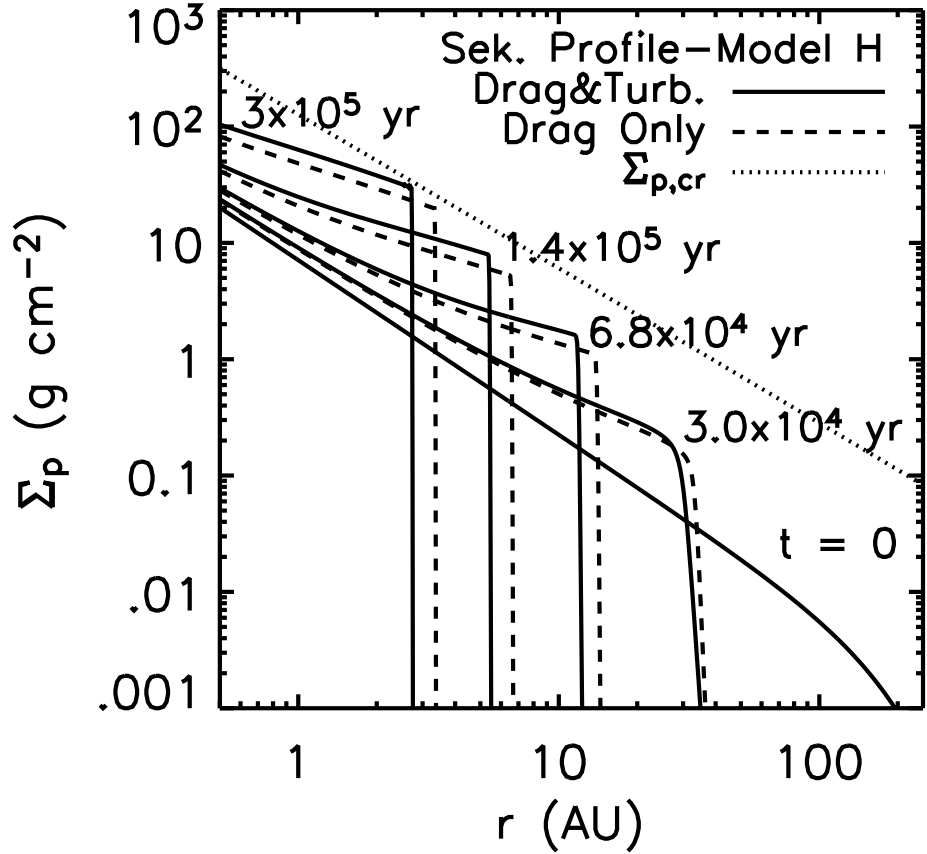}{305pt}{0}{80}{80}{-210}{-10}
\caption{Evolution of the surface density of solids with time.
Solids are modelled as millimeter-sized compact spheres. The particle
disk shrinks and amplifies its surface density as it accretes
inward by aerodynamic drag. Solid lines
account for angular momentum transport by
Kelvin-Helmholtz turbulence, which is seen to hasten the pile-up.
At $t = 3 \times 10^5$ yr, the surface density at $r \sim 3$ AU
exceeds the critical threshold (dotted line) above which gravitational
instability is thought to occur. Figure taken from Youdin \& Chiang (2003).}
\end{figure}

Future work should follow the evolution of self-gravitating sheets
of dust and calculate the size spectrum of resultant planetesimals.

\subsection{A Cautionary Remark}
Astrophysicists may wax eloquent on grain growth by gas-phase condensation,
grain-grain collisions, and gravitational instability, but until they
understand how chondritic meteorites are formed, it is possible that
they are missing the boat. Chondritic meteorites are the oldest creations
of the solar system; when we refer to the age of the solar
system of 4.566 billion years, we are referring to the lead-lead ages
of such meteorites. What has confounded meteoriticists and astrophysicists
alike is why they contain nearly identically sized, once molten
spheres, a.k.a {\it chondrules}, and their more refractory
counterparts, {\it calcium-aluminum inclusions (CAIs)}.
Their record-breaking ages, nearly solar composition,
and enormous volume-filling fraction (50\%--90\% of the host
meteorite) all suggest that these igneous, mm-to-cm-sized marbles
are the building blocks of planetesimals; that to traverse the size ladder
from microns to meters, one must necessarily melt and agglomerate
chondrules first.
Recently, Desch \& Connolly (2002) have shown that the thermal
histories of chondrules can be reproduced by
processing of solid particles through strong hydrodynamic
shocks having Mach numbers of 5--10 in the (possibly metallicity
enhanced) solar nebula. The origin of such shocks, and the mechanism
by which heated chondrules are collected into their host bodies
with the observed high efficiencies, are unknown; see Chiang (2002)
for a short review. Shocks are naturally generated by non-linear
steepening of turbulent fluctuations
within gravitationally unstable (Gammie 2001; Johnson \& Gammie 2003)
or magneto-rotationally unstable disks, though whether
the requisite Mach numbers can be attained is unclear.

\section{Dusty Signposts of Planets}
Young planets can leave their mark on the reservoirs of
dust from which they arose by gravitationally torquing
dust into signature non-axisymmetric patterns.
The phenomenon is well studied in the solar system context;
for example, Kuiper belt objects (read: very large dust particles)
that inhabit the exterior 3:2 mean-motion resonance with Neptune
preferentially attain perihelia at longitudes displaced
$\pm$$90\deg$ away from that planet. An instantaneous
snapshot of thousands of 3:2 resonant objects would reveal
a characteristic ``keyhole'' pattern that rotates with the
angular speed of Neptune; see Figure 5, taken from Chiang \& Jordan (2002).

\begin{figure}
\plotfiddle{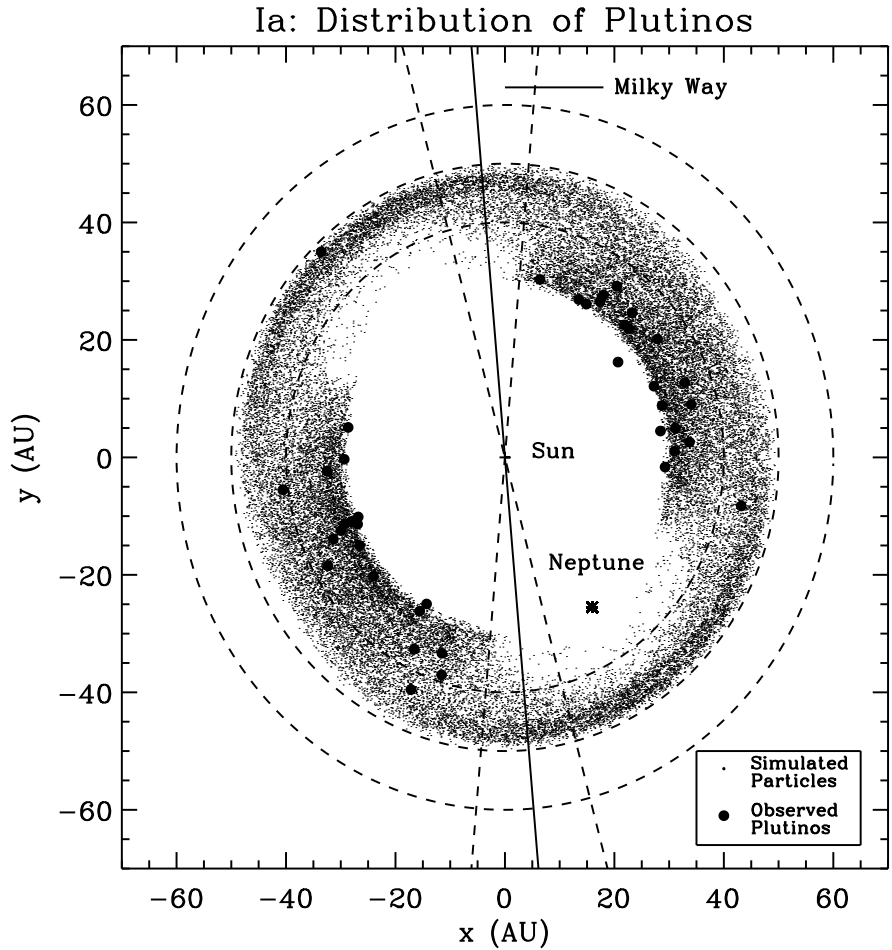}{285pt}{0}{65}{65}{-190}{-10}
\caption{Theoretical snapshot, viewed above the plane of the
solar system, of the spatial distribution of Kuiper belt objects
that are 3:2 resonant with Neptune. The clumping of objects
$\pm$$90\deg$ from Neptune's position follows from the shape
of the resonant perturbation potential established by
Neptune, not from the self-gravity of Kuiper belt objects.
The entire keyhole-like pattern rotates with the
angular speed of Neptune. Dashed circles indicate
heliocentric distance of 40, 50, and 60 AU. Solid circles
denote positions of known 3:2 resonant objects (also called
``Plutinos.'') Similar patterns
might be traced out by dust particles in resonance
with extrasolar planets. Figure taken from Chiang \& Jordan (2002).}
\end{figure}

Observations of similar non-axisymmetric patterns of dust
surrounding
other stars are used to implicate the existence
of planets. Thermal emission at sub-mm wavelengths
is spatially resolved around the stars $\epsilon$ Eridani
(Ozernoy et al.~2000), Fomalhaut (Holland et al.~2003),
and Vega (Wilner et al.~2002).
Around each star are several clumps of emission interpreted
to arise from dust particles trapped in mean-motion resonances
with an as yet unseen planet. The dust particles are hypothesized
to be generated from the collisions of yet larger planetesimals;
Poynting-Robertson drag brings the dust into mean-motion
resonance, where they resonantly librate for a spell,
and eventually brings them out again. In the case of Vega,
two clumps of emission are detected at a projected
stellocentric distance of $\sim$80 AU, and appear nearly diametrically
opposed. Ascribing these clumps to be the limb-brightened
edges of a circumstellar ring viewed edge-on is not the most
natural interpretation, since Vega the star is thought to be
viewed nearly pole-on (Gulliver et al.~1994). The clumps
are alternatively interpreted as dust particles temporarily
trapped in $n$:1 resonance with a Jovian-mass planet on a highly
eccentric orbit having a semi-major axis of 40 AU,
where $n$ can be an integer greater than 3
(Wilner et al.~2002; Kuchner \& Holman 2002).
While the parameters
of the planet---e.g., the mass and orbital eccentricity---can not
be pinned down with great accuracy, and while there is concern
as to whether interparticle collisions might vaporize dust en route
to resonance trapping, the scenario at least has a robust, testable
prediction: the clumps should rotate on the sky
with a pattern speed that is not equal
to the local Keplerian frequency, but is rather equal to half
the mean orbital frequency of the planet (Kuchner \& Holman 2002).
The magnitude of the expected pattern speed is sufficiently
large in the case of Vega---$1\deg$/yr---that multi-epoch
observations should be able to test this prediction.

\acknowledgements
Support for this work was provided by Hubble Space Telescope
Theory grant HST-AR-09514.01-A. We thank the organizers
for a highly instructive conference.

\end{document}